\documentclass[a4paper,fleqn,usenatbib]{mnras}

\usepackage[T1]{fontenc}
\usepackage{ae,aecompl}

\usepackage{graphicx}	
\usepackage{amsmath}	
\usepackage{amssymb}	
\usepackage{color}

\definecolor{grey}{rgb}{0.7,0.7,0.7}

\newcommand{\bluetides}{{\sc BlueTides}}
\newcommand{\jwst}{{\em JWST}}

\title[Dust Obscured Star Forming Galaxies at $z>8$]{Dust Obscured Star Forming Galaxies in the Early Universe}

\author[Stephen M. Wilkins et al.]{
Stephen M. Wilkins,$^{1}$\thanks{E-mail: s.wilkins@sussex.ac.uk}
Yu Feng,$^{2,3}$ 
Tiziana Di Matteo,$^{2}$ 
Rupert Croft,$^{2}$\newauthor 
Christopher C. Lovell,$^{1}$ 
Peter Thomas,$^{1}$
\\
$^1$\,Astronomy Centre, Department of Physics and Astronomy, University of Sussex, Brighton, BN1 9QH, UK \\
$^2$\,McWilliams Center for Cosmology, Carnegie Mellon University, Pittsburgh PA, 15213, USA \\
$^3$\,Berkeley Center for Cosmological Physics, University of California, Berkeley, Berkeley CA, 94720, USA \\
}

\date{Accepted XXX. Received YYY; in original form ZZZ}

\pubyear{2017}

\begin{document}
\label{firstpage}
\pagerange{\pageref{firstpage}--\pageref{lastpage}}
\maketitle


\begin{abstract}

Motivated by recent observational constraints on dust reprocessed emission in star forming galaxies at $z\sim 6$ and above we use the very-large cosmological hydrodynamical simulation \bluetides\ to explore predictions for the amount of dust obscured star formation in the early Universe ($z>8$). \bluetides\ matches current observational constraints on both the UV luminosity function and galaxy stellar mass function and predicts that approximately $90\%$ of the star formation in high-mass ($M_{*}>10^{10}\,{\rm M_{\odot}}$) galaxies at $z=8$ is already obscured by dust. The relationship between dust attenuation and stellar mass predicted by \bluetides\ is consistent with that observed at lower redshift. However, observations of several individual objects at $z>6$ are discrepant with the predictions, though it is possible their uncertainties may have been underestimated. We find that the predicted surface density of $z\ge 8$ sub-mm sources is below that accessible to current {\em Herschel}, SCUBA-2, and ALMA sub-mm surveys. However, as ALMA continues to accrue additional surface area the population of $z>8$ dust-obscured galaxies may become accessible in the near future.

\end{abstract}


\section{Introduction}

The {\em Hubble Space Telescope} has been enormously successful in extending our knowledge of the galaxy population to the early Universe, with $\sim 1000$ galaxies now identified above $z\sim 6$ \citep[e.g.][]{Bouwens2015a, Finkelstein2015} and the first, albeit small, samples identified at $z>10$ \citep[e.g.][]{Bouwens2015a, Oesch2015b, Oesch2016}. With the imminent launch of the {\em James Webb Space Telescope (JWST)} this frontier will expand further with potentially hundreds of galaxies discovered at $z>10$ and the first samples discovered at $z>12$. 

Despite the success of {\em Hubble} in identifying galaxies in the early Universe it is only capable of probing the rest-frame UV emission of these galaxies and as such is limited to probing unobscured star formation. While it may be possible to constrain dust attenuation using the observed rest-frame UV continuum slope \citep[e.g.][]{Meurer1999, Wilkins2011b, Wilkins2016a, Bouwens2012b, Bouwens2014a} this is sensitive to a range of assumptions (including the intrinsic UV continuum slope, the dust attenuation curve, and the geometry) that need to be made \citep[e.g.][]{Wilkins2012b, Wilkins2013b, Wilkins2016a}. Indeed, the applicability of $z\sim 0$ relations (the IRX-$\beta$ relationship) have been challenged by recent observational constraints from the Atacama Large Millimetre Array (ALMA) \citep[e.g.][]{Bouwens2016_ASPECS}. 

The key to constraining obscured star formation is to measure the dust-reprocessed rest-frame UV/optical emission in the rest-frame IR. While single-dish far-IR/sub-mm observatories such as {\em Herschel} and SCUBA-2 offer the wavelength coverage capable of probing the rest-frame IR emission at high-redshift they lack the sensitivity to detect all but the most intensely star forming galaxies at high-redshift and are susceptible to considerable confusion. For example, the SCUBA-2 Cosmology Legacy Survey \citep[S2CLS,][]{Geach2016} is only capable of detecting individual sources at $z\sim 8$ with intrinsic\footnote{It is of course possible to probe lower SFRs with strongly lensed sources but these are relatively rare.} star formation rates of several hundred ${\rm M_{\odot} yr^{-1}}$ while for the {\em Herschel} multi-tiered extragalactic survey \citep[HerMES,][]{Oliver2012} the threshold is $>1000\,{\rm M_{\odot} yr^{-1}}$. However, it is now possible to efficiently probe obscured star formation rates of $\sim 10\,{\rm M_{\odot} yr^{-1}}$ in individual galaxies using ALMA. While at present there are only four detections \citep[e.g.][]{Watson2015, Willott2015, Laporte2017} and a handful of deep non-detections \citep[e.g.][]{Ouchi2013, Schaerer2015} at $z>6$, these are expected to grow rapidly.

While powerful, ALMA follow-up of sources detected by {\em Hubble} is biased to galaxies with relatively low levels of obscuration and may not provide the full picture of dust obscured star formation at high-redshift. This is a consequence of the fact that heavily obscured galaxies will not only have reduced rest-frame UV luminosities but will also have red rest-frame UV continuum colours that may push them out of the Lyman-break selection window. Identifying the population of heavily obscured galaxies at high-redshift then relies on blank field sub-mm surveys\footnote{It may however also be possible to use {\em JWST}/MIRI to perform a rest-frame near-IR (which is less affected by dust) selection of galaxies at $z>6$ combined with ALMA follow-up.} such as the SCUBA-2 Cosmology Legacy Survey, or {\em Herschel} multi-tiered extragalactic survey. However, because of their sensitivity these surveys have, thus far, only revealed a single object at $z>6$, the {\em Herschel} selected lensed galaxy HFLS3 \citep{Riechers2013, Cooray2014}. More recently ALMA has begun blank field surveys, with both the ALMA Spectroscopic Survey in the Hubble Ultra Deep Field \citep[ASPECS,][]{Walter2016} and the \citet{Dunlop2016} survey surveying the {\em Hubble} Ultra Deep Field. While reaching much higher sensitivities these surveys have, thus far, only probed relative small areas ($\ll 1\,{\rm deg^{2}}$) and have not yet yielded any dust-obscured high-redshift galaxies.

Concurrent with our growing ability to observationally explore the early Universe, galaxy formation modelling has also improved dramatically. Most recently, it is now possible to perform high-resolution hydrodynamical simulations following which self consistently follow the evolution of dark matter and baryons in cosmologically representative volumes \citep[e.g.][]{DiMatteo2012, Vogelsberger2014, Schaye2015, Khandai2015}.

In this study we use the large ($(400/h\approx 577)^{3}\,{\rm cMpc^3}$) cosmological hydrodynamical simulation \bluetides\ \citep{Feng2015,Feng2016,Wilkins2017a} to investigate dust obscured star formation in the early Universe ($z>8$). We begin in Section \ref{sec:BT} by briefly describing the simulation and how we model the attenuation by dust. In Section \ref{sec:fesc} we present predictions for the far-UV attenuation as a function of stellar mass and compare this to observational constraints at both intermediate and high-redshift. Following this, in Section \ref{sec:sfr}, we present predictions for the intrinsic, obscured, and unobscured star formation rate (SFR) distribution functions. In Section \ref{sec:surface_densities} we model the rest-frame IR luminosities of galaxies at $z\sim 8$ and calculate the predicted surface density of sources at $z>8$. Finally, in Section \ref{sec:c} we present our conclusions.

\section{The BlueTides Simulation}\label{sec:BT}

The \bluetides\ simulation \citep[\url{http://bluetides-project.org/}, see][for description of the simulation physics]{Feng2015,Feng2016} is an extremely large galaxy formation simulation carried out using the smoothed particle hydrodynamics (SPH) code {\sc MP-Gadget}. Phase I of \bluetides\ evolved a $(400/h\approx 577)^{3}\,{\rm cMpc^3}$ cube to $z=8$ using $2\,\times\, 7040^{3}$ particles to $z=8$. The simulation was run assuming the {\em Wilkinson Microwave Anisotropy Probe} nine year data release \citep{Hinshaw2013}. 

By $z=8$ there are approximately 200 million objects identified within the simulation volume and, of these, almost 160,000 have stellar masses greater than $10^{8}\,{\rm M_{\odot}}$. The properties of galaxies in the simulation are extensively described in \citet{Feng2015,Feng2016, Wilkins2016b, Wilkins2016c, Waters2016a, Waters2016b, DiMatteo2016, Wilkins2017a}.

\subsection{Modelling attenuation by dust}

To estimate the dust attenuation of galaxies in \bluetides\ we employ a simple scheme which links the smoothed metal density integrated along lines of sight to each star particle within each galaxy to the dust optical depth in the $V$-band ($550\,{\rm nm}$) $\tau_V$ with attenuation at other wavelengths determined assuming a simple attenuation curve of the form,
\begin{equation}
\tau_{\lambda} = \tau_V\times (\lambda/550{\rm nm})^{-1}.
\end{equation}
This model has a single free parameter $\kappa$ which effectively links the surface density of metals to the optical depth (and thus surface density of metals). This parameter $\kappa$ is fit to recover the shape of the of the observed $z=8$ far-UV luminosity function. For a full description of this model, see \citet{Wilkins2017a}. 

As discussed in \citet{Wilkins2017a} it is important to note that this model is a simplification. While the quantities of dust and metals are expected to be linked, because of the different formation mechanisms, they are not expected to trace each other exactly, with the ratio varying from galaxy-to-galaxy.

It is also important to note that while the effect of dust on the spectral energy distribution is, in general, sensitive to the slope of the attenuation curve (assumed to be $\gamma=-1$ in this analysis) the energy absorbed and re-emitted in the IR is only weakly sensitive ($<20\%$) to the choice of curve. This is because a change to the slope $\gamma$ changes the best-fit value of $\kappa$.


\section{Far-UV escape fraction}\label{sec:fesc}

A key observational property of galaxies is how the fraction of obscured (or unobscured) star formation varies with properties like the stellar mass or far-UV luminosity. This can alternatively be expressed in terms of the far-UV attenuation $A_{\rm FUV}$ or far-UV escape fraction\footnote{This is different from the Lyman-Continuum (or ionising photon) escape fraction which is driven mostly by the density and distribution of hydrogen surrounding ionising sources.} $A_{\rm FUV} = -2.5\log_{10}(f_{\rm esc, FUV})$. At intermediate-redshift ($z\sim 1-4$) the far-UV escape fraction is observed to decline rapidly with increasing stellar mass \citep[e.g.][]{Heinis2014, Pannella2015}, at least at stellar masses $>10^{10}\,{\rm M_{\odot}}$.

\begin{figure}
\centering
\includegraphics[width=20pc]{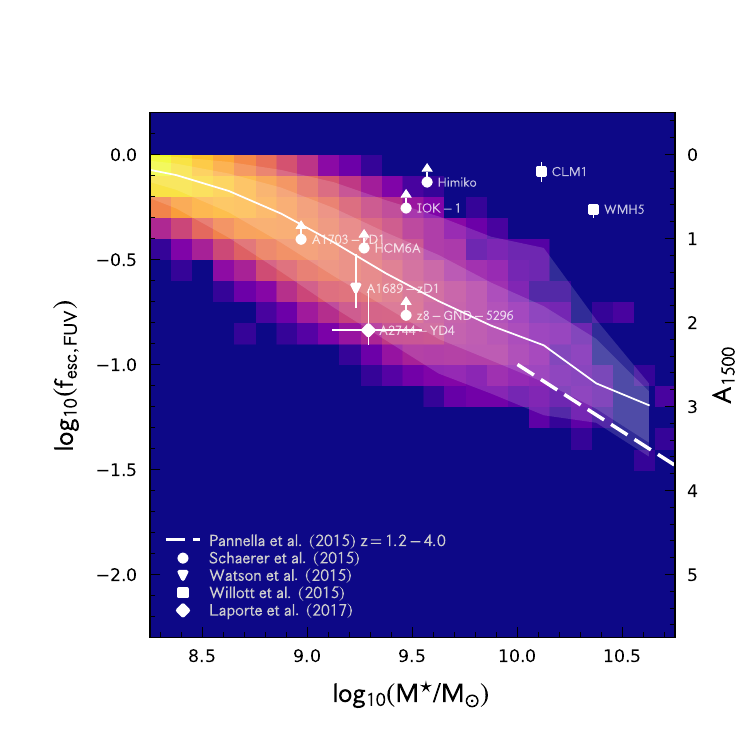}
\caption{The distribution of far-UV escape fractions and stellar masses predicted by the \bluetides\ simulation at $z=8$. The shading denotes the logarithmic density of galaxies. The solid line shows the median escape fraction in bins of $\log_{10}(M_{*}/M_{\odot})$. The inner and outer shaded regions show the range of containing the central $68.3\%$ and $95.4\%$ of objects respectively. Observational constraints at $z>6$ from \citet{Willott2015}, \citet{Schaerer2015}, \citet{Watson2015}, and  \citet{Laporte2017} are also shown alongside observational constraints at $z=1.2-4.0$ from \citet{Pannella2015}. The object HFLS3 \citep{Riechers2013, Cooray2014}, which has $\log_{10}(f_{\rm esc, FUV})\approx -4$, is omitted as it falls far below other observations. Observed stellar masses are converted to assume a \citet{Chabrier2003} initial mass function. The right-hand axes shows the corresponding far-UV attenuation in magnitudes. Tabulated values of $f_{\rm esc, FUV}$ in bins of stellar mass are presented in Table \ref{tab:L_fesc}.}
\label{fig:L_fesc}
\end{figure}

\begin{figure}
\centering
\includegraphics[width=20pc]{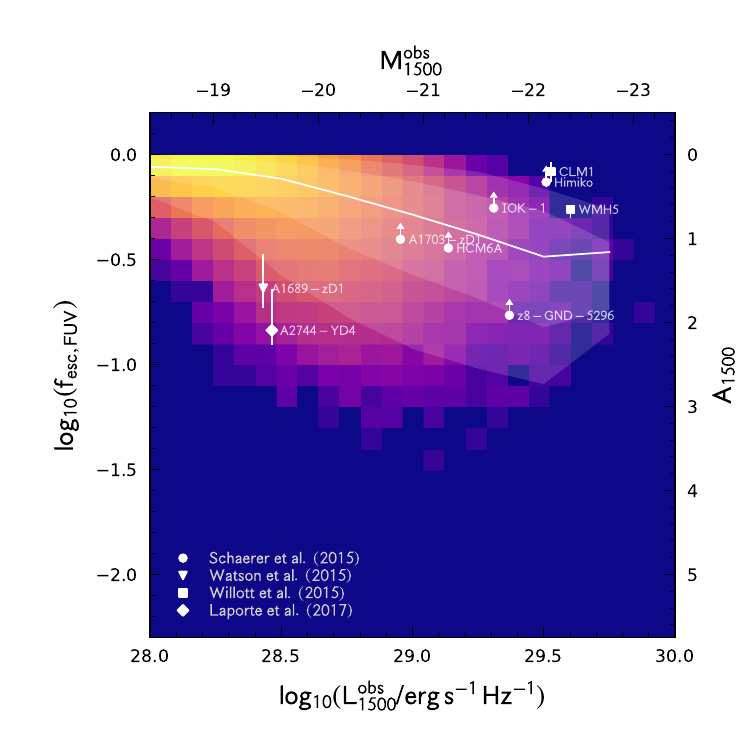}
\caption{Similar to Figure \ref{fig:L_fesc} but instead showing the trend with observed rest-frame UV luminosity. Tabulated values of $f_{\rm esc, FUV}$ in bins of observed UV luminosity are presented in Table \ref{tab:L_fesc_L1500}.}
\label{fig:L_fesc_L1500}
\end{figure}

The far-UV escape fraction as a function of the stellar mass and observed rest-frame UV luminosity predicted by \bluetides\ at $z=8$ are shown in Figures \ref{fig:L_fesc} and \ref{fig:L_fesc_L1500} respectively. As shown in Figure \ref{fig:L_fesc} the predicted far-UV escape fraction declines steadily with stellar mass from $f_{\rm esc, FUV}\approx 0.7$ at $M_{*}=10^{8.5}\,{\rm M_{\odot}}$ to $f_{\rm esc, FUV}\approx 0.1$ ($A_{1500}\approx 2.5$) at $M_{*}=10^{10}\,{\rm M_{\odot}}$. The relationship it tight with predicted scatter is approximately $0.1-0.2\,{\rm dex}$. At stellar masses above $10^{10.4}\,{\rm M_{\odot}}$ there are fewer than 10 objects in each bin and it becomes impossible to reliably constrain the relationship. The average escape fraction also decreases with observed UV luminosity though it is less pronounced than the trend with stellar mass with more scatter ($0.2-0.3\,{\rm dex}$). This simply reflects that massive intrinsically bright galaxies are pushed to lower observed UV luminosities by increased dust attenuation.

Key to confirming these predictions are observations permitting the measurement of the stellar masses, and both obscured (as probed by the rest-frame IR luminosity) and unobscured (i.e. far-UV inferred) star formation rates. While at present there are only weak constraints at $z>8$ there are now a handful of objects with the requisite observations at $z>6$.  

At present there is only a single sub-mm selected galaxy at $z>6$, HFLS3 \citep{Riechers2013, Cooray2014}. HFLS3 has an inferred stellar mass \citep[$\sim 5\times 10^{10}\,{\rm M_{\odot}}$][]{Cooray2014} outside the range accessible to \bluetides\ at $z=8$. While the inferred far-UV escape fraction of $\log_{10}(f_{\rm esc, FUV})\approx -4$ lies well below the extrapolation of the trend at lower-masses the difference in redshift and stellar mass complicate using HFLS3 as a robust constraint on the model. It is also interesting to note that the relationship between stellar mass and far-UV escape fraction predicted by \bluetides\ overlaps that found by \citet{Pannella2015} at $z=1.2-4$ and \citet{Heinis2014} at $z=1.5-4$.

While there remains limited samples of IR detected galaxies at high-redsift there now exists a small sample of UV selected Lyman-break galaxies (LBGs) at $z>6$ with sub-mm flux detections (or strong upper limits) from ALMA or the Plateau de Bure interferometer (PdBI) \citep{Schaerer2015, Watson2015, Willott2015, Laporte2017}. These results are shown in both Figures \ref{fig:L_fesc} and \ref{fig:L_fesc_L1500}. In each case the far-UV escape fraction is estimated from the reported IR and total or far-UV SFR, i.e. $f_{\rm esc, FUV}={\rm SFR_{UV}}/{\rm SFR_{total}}={\rm SFR_{FUV}}/({\rm SFR_{IR}+SFR_{FUV}})$. The measured UV luminosities, stellar masses, and escape fractions of the lensed LBG A1689-zD1 \citep[][$z\approx 7.5$]{Watson2015} and \citep[][$z\approx 8.4$]{Laporte2017} are consistent with both sets of predictions from \bluetides. While the objects studied by \citet{Schaerer2015} are all undetected by ALMA or PdBI they provide strong constraints on the far-UV escape fraction. Three of these have lower-limits consistent with the trend with stellar mass predicted by \bluetides\ while two (IOK-1 and Himiko) appear to have escape fractions outside the central $95\%$ range for their stellar masses. When comparing the observed UV luminosity only Himiko lies outside the central $95\%$ range, and then only by a small amount. Both objects (CLM1 and WMH5) studied by \citet{Willott2015} lie significantly outside central $95\%$ range based on their reported stellar masses, and UV/IR inferred SFRs. However, when comparing with observed UV luminosity only CLM1 lies outside the central $95\%$ range, and then only by a small amount.

This comparison reveals some significant tension between current observational results and the simulation predictions, particularly when using stellar masses. However, as the observed galaxies are predominantly (all with the exception of HFLS3) selected using the Lyman-break technique (or an effective variant thereof) they are biased to being both UV bright and having relatively low dust attenuation. As such it is unsurprising that this observed population is observed to have lower dust attenuation than the entire population. While this selection bias alone is capable of explaining the tension in the trend with observed UV luminosity it is incapable of fully explaining the trend with stellar mass. However, there are several potential solutions to this remaining discrepancy: firstly, the stellar masses of CLM1, WMH5, and to a lesser extent Himiko and IOK-1 may have been overestimated possibly due to the lack of high quality rest-frame optical/near-IR observations and the complications of strong nebular emission. In the near future, improved rest-frame optical photometry and spectroscopy from \jwst\ should provide much stronger constraints on the stellar masses. Secondly, the inferred IR luminosities (and thus the IR inferred SFRs and far-UV escape fractions) may have been underestimated, perhaps by assuming an incorrect far-IR SED template. The values reported by \citet{Willott2015}, for example, assume a simple greybody SED with $\beta=1.6$ and $T=30\,{\rm K}$. If instead a temperature of $T=45\,{\rm K}$ is assumed the inferred IR luminosity increases by more than a factor of 3. If instead a combination of a mid-IR power-law and greybody \citep[i.e.][]{Casey2012} with $\beta=1.6$, $T=45\,{\rm K}$, and $\alpha=2.0$ is assumed the inferred IR luminosity would increase by a factor of $\sim 5$ compared to the original value. In either case this would be sufficient to leave the far-UV escape fractions consistent with the predictions from \bluetides. Key to overcoming this limitation are observations in multiple far-IR/sub-mm bands providing the ability to constrain the shape of the sub-mm/IR SED and thus the total IR luminosity or simply fitting assuming a range of temperatures and other parameters to generate releastic uncertainties. Thirdly, the deviation of these objects from the \bluetides\ predictions may be a consequence of our simple dust attenuation model. As noted in the preceding section and discussed in \citet{Wilkins2017a} it is unlikely that the dust-to-metal ratio is uniform (as assumed here) but instead is sensitive to the assembly history of the galaxy and other factors. A more sophisticated model may result in a wider spread of escape fractions at a given stellar mass possibly bringing the model predictions inline with the current observational constraints while also maintaining the good agreement with the observed UV luminosity function. Finally, this discrepancy may hint at a deeper issue with the physics implemented in the simulation.


\section{Star Formation Rate Distribution Functions}\label{sec:sfr}

Using the far-UV escape fractions we can approximately split the predicted intrinsic (or total) SFR distribution function into obscured (or infrared inferred, ${\rm SFR_{IR}}=(1-f_{\rm esc, FUV})\times {\rm SFR_{tot}}$) and unobscured (or far-UV inferred, ${\rm SFR_{FUV}}=f_{\rm esc, FUV}\times {\rm SFR_{tot}}$) SFR distribution functions. 

It is worth noting that this definition differs slightly from that used observationally. Observationally, the obscured and unobscured SFRs are obtained from combining total IR and observed UV luminosities with a theoretically motivated calibration \citep[e.g.][]{KE12}. As noted by \citet{Wilkins2016c} and \citet{Wilkins2012b} these calibrations may not be appropriate for certain populations of galaxies. A further complicating factor is that within the simulation the SFR can be extracted using two different approaches: using the instantaneous SFR of the gas particles or by using the number of star particles formed averaged over some timescale. We choose to define the total SFR as the average SF activity over the last $100\,{\rm Myr}$. This yields SFRs $\sim 0.1\,{\rm dex}$ smaller than those based on the instantaneous gas SFRs and is more closely comparable to SFRs inferred observationally from the UV/IR than those based on the instantaneous gas properties or a shorter timescale (though those are more suitable when comparing SFRs inferred from recombination lines).

\begin{figure*}
\centering
\includegraphics[width=40pc]{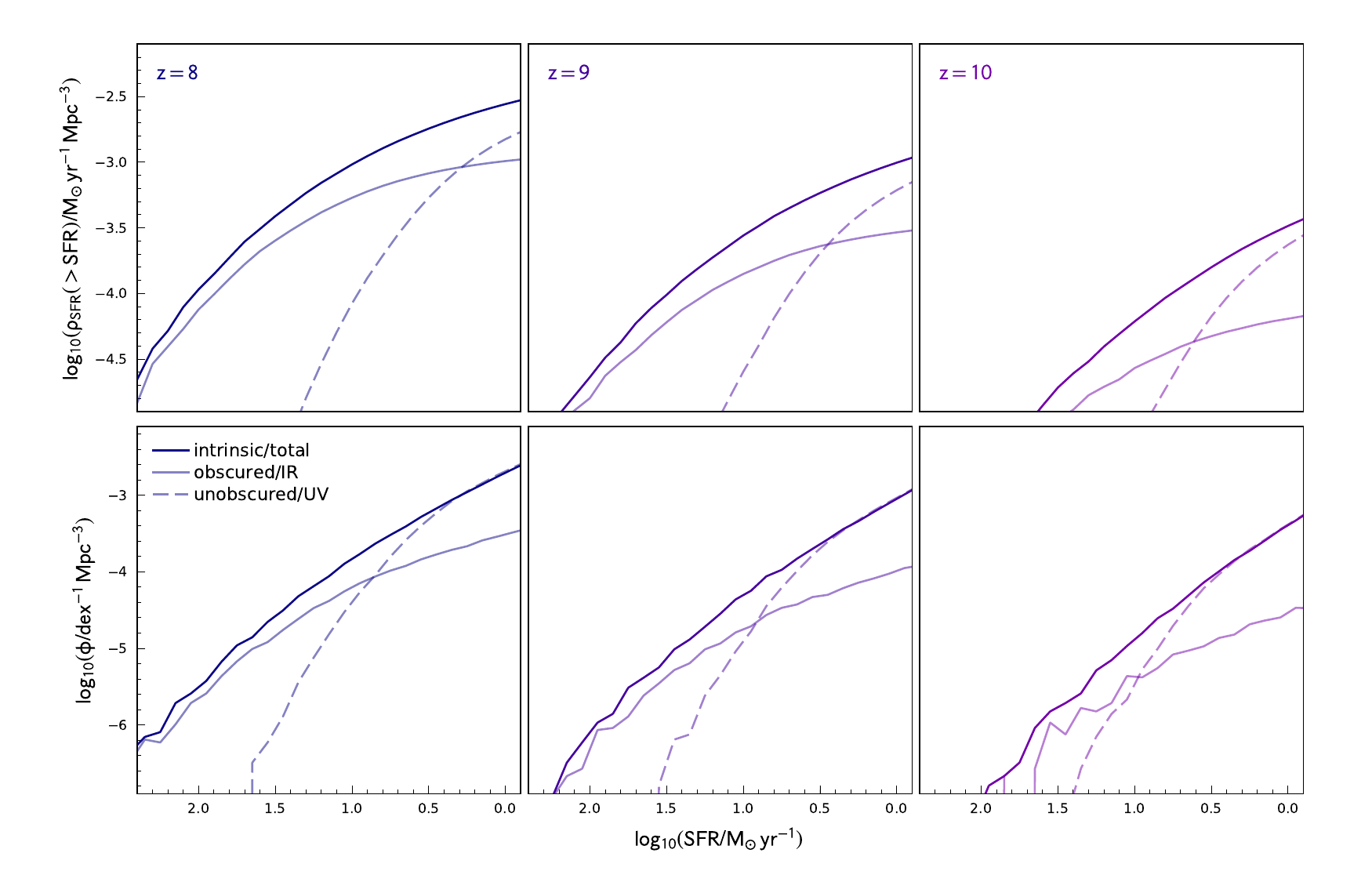}
\caption{Intrinsic/total (solid dark line), obscured (solid light line), and unobscured (dashed line) star formation rate distribution functions (bottom panels) and cumulative star formation rate densities (top panels) predicted by \bluetides\ at $z\in\{8,9,10\}$.}
\label{fig:SFRDF_multi}
\end{figure*}

The intrinsic/total, obscured, and unobscured SFR distribution functions for $z\in\{8,9,10\}$ are shown in Figure \ref{fig:SFRDF_multi} alongside the cumulative star formation rate density. Obscured star formation dominates the SFR distribution function at ${\rm SFR}>10\,{\rm M_{\odot} yr^{-1}}$ (at all redshifts) and dominates the cumulative star formation rate density at ${\rm SFR}>2\,{\rm M_{\odot} yr^{-1}}$ at $z=8$. At higher-redshift, obscured star formation dominates the cumulative SFR density at higher SFRs due to decreasing contribution of high-SFR to the total SFR density.

\section{Sub-mm Fluxes and Surface Densities}\label{sec:surface_densities}

We next make predictions for the surface density of sub-mm sources powered exclusively by dust reprocessed starlight. In this analysis we omit the contribution of AGN as a source of heating. Approximately $5\%$ of the intrinsic UV emission is predicted to produced by AGN in galaxies with $\log_{10}(M_{*}/{\rm M_{\odot}})=9-10$ \citep[see][]{Wilkins2017a}. At $\log_{10}(M_{*}/{\rm M_{\odot}})>10$ the average contribution is similar, however, there is a higher proportion ($\approx 25\%$) of galaxies where the AGN contributes $>10\%$ of the intrinsic luminosity, including 6 objects which are AGN dominated. As such, the contribution of AGN to the sub-mm number counts at $z>8$ is expected to be small except at the extreme fluxes. To make these predictions we self-consistently model the sub-mm fluxes by redistributing the energy absorbed by dust in the UV/optical into the IR assuming a model IR SED. We consider two simple parameterisations for the IR SED: a simple greybody (with the emissivity fixed to $\beta = 1.6$) and the \citet{Casey2012} greybody + mid-IR power-law SED (with emissivity $\beta = 1.6$, and mid-IR power-law index $\alpha=2.0$).

\begin{figure*}
\centering
\includegraphics[width=40pc]{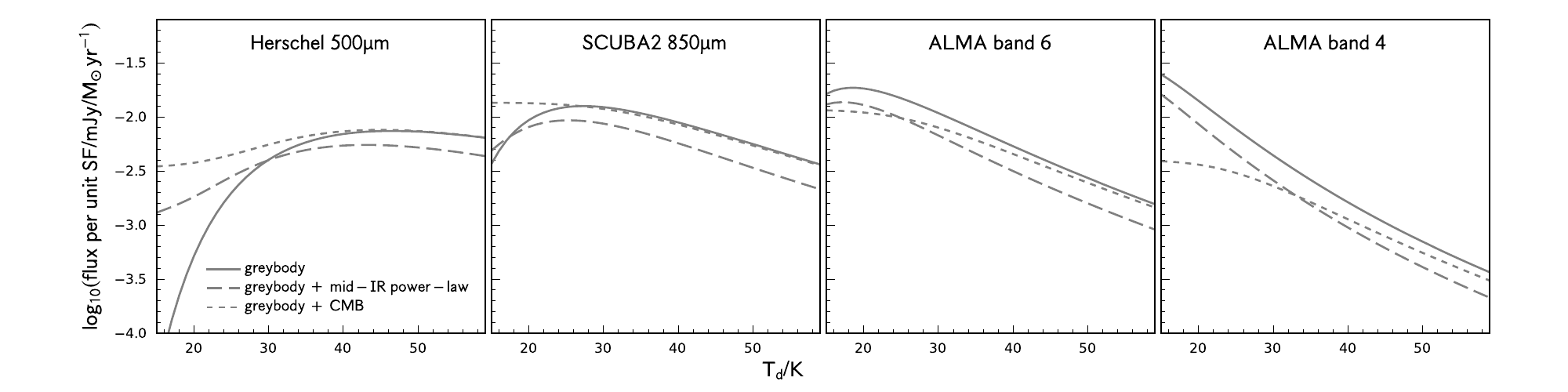}
\caption{The expected flux per unit obscured star formation for a galaxy at $z=8$ as a function of dust temperature for 3 bands for both a greybody ($\beta = 1.6$) SED and a greybody with mid-IR power-law SED \citep[][assuming $\beta = 1.6$, $\alpha=2.0$]{Casey2012}. The dotted line shows the prediction when both heating by the CMB and the observability against the CMB is included for the greybody SED.}
\label{fig:flux_T}
\end{figure*}

The choice of template can have a significant effect on the predicted sub-mm fluxes. As an illustration of the effect of the far-IR SED on the inferred sub-mm fluxes in Figure \ref{fig:flux_T} we show how the sub-mm flux per unit obscured star formation (assuming a model with a simple constant SFH) for a galaxy at $z=8$ varies with the choice of wavelength, SED model, dust temperature, and the impact of the Cosmic Microwave Background (CMB). The impact of the CMB is modelled using the formalism described by \citet{daCunha2013} in which the CMB provides both an additional source of heating and a background against which emission is measured. 

As can be seen in Figure \ref{fig:flux_T} observations using ALMA bands 4 and 6 are strongly sensitive to the choice of model and dust temperature with the expected flux varying by approximately an order of magnitude for $T=20\to 50\,{\rm K}$. Varying the other SED model parameters (the emissivity and the near-IR power-law index) further increases the range of possible flux values. The impact of the CMB is also significant at lower temperatures ($T_{d}<35\,{\rm K}$) increasing the observed flux in the shorter wavelength bands and decreasing it at longer-wavelengths. 

It is also worth noting that low dust temperatures ($T_{d}<25\,{\rm K}$) are difficult to obtain in the simulation. This is because the inferred mass dust mass \citep[following][]{Casey2012} is comparable to or exceeds the mass available in metals.

\begin{figure*}
\centering
\includegraphics[width=40pc]{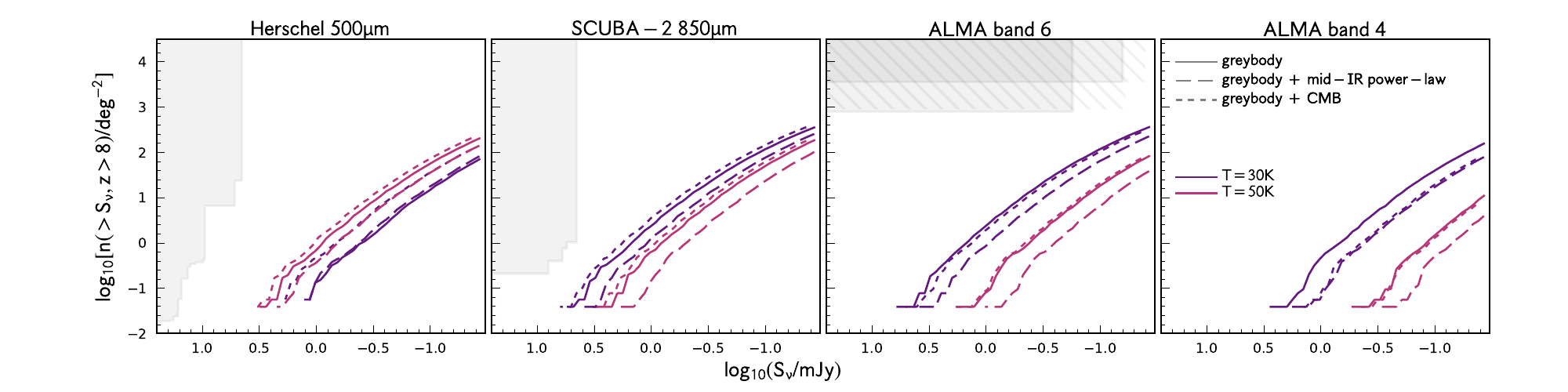}
\caption{The predicted cumulative surface density of sources at $z>8$ in the {\em Herschel}/SPIRE $500\mu m$ band, the SCUBA-2 $850\mu m$ band, and ALMA band 6. Predictions are made assuming both a simple greybody ($T/{\rm K}\in\{30,50\}$, $\beta = 1.6$) and a greybody with mid-IR power-law ($T/{\rm K}\in\{30,50\}$, $\beta = 1.6$, $\alpha=2.0$). Shaded areas show the region probed by various surveys. For the {\em Herschel}/SPIRE $500\mu m$ band this includes the combination of the various {\em Herschel} surveys \citep{Oliver2012}, for SCUBA-2 $850\mu m$ band the SCUBA-2 cosmology legacy survey \citep{Geach2016}, and for ALMA band 6 the ALMA Spectroscopic Survey in the Hubble Ultra Deep Field \citep[ASPECS,][]{Walter2016} (wider/shallower region) and the \citet{Dunlop2016} ALMA imaging of the {\em Hubble} Ultra Deep Field (narrower/deeper region). The planned extension to the ASPECS survey is shown by the hatched region.}
\label{fig:culmulative_surface_density_multi}
\end{figure*}

The resulting predicted cumulative surface densities of sources at $z>8$ are shown in Figure \ref{fig:culmulative_surface_density_multi} for the {\em Herschel}/SPIRE $500\mu m$ band, the SCUBA-2 $850\mu m$ band, and ALMA bands 6 and 4. Predictions are made assuming both a simple greybody ($T/{\rm K}\in\{30,50\}$, $\beta = 1.6$) and a greybody with mid-IR power-law ($T/{\rm K}\in\{30,50\}$, $\beta = 1.6$, $\alpha=2.0$). Including the effect of the CMB on the {\em Herschel}/SPIRE $500\mu m$ band increases the fluxes by $<0.1\,{\rm dex}$ assuming $T=30\,{\rm K}$ while in the other bands the result lies between the two SED models.

We also show the area probed by various surveys as a function of the $5\sigma$ sensitivity. These include various {\em Herschel} surveys \citep{Oliver2012}, the SCUBA-2 cosmology legacy survey \citep{Geach2016}, the ALMA Spectroscopic Survey in the Hubble Ultra Deep Field \citep[ASPECS,][]{Walter2016} and the \citet{Dunlop2016} ALMA imaging of the {\em Hubble} Ultra Deep Field. The predicted surface densities all lie well outside the current survey limits, even for an optimistic scenario of relatively cool dust. As such our predictions are consistent with the absence of any individual galaxies at $z>8$ found as yet in the individual surveys. 

While current sub-mm surveys either lack the area or sensitivity to detect galaxies at $z>8$ it is likely that ALMA will {\em eventually} build up sufficient area to either detect or place strong constraints on the numbers of obscured galaxies at $z>8$. This will provide a strong constraint on the physics of massive galaxy formation in the early Universe.

\section{Conclusions}\label{sec:c}

The very-large cosmological hydrodynamical simulation \bluetides\ is capable of reproducing current observational constraints on both the UV luminosity function and galaxy stellar mass function at $z\ge 8$. However, it also predicts that many massive $z\sim 8$ galaxies are already heavily obscured by dust. We have compared predictions from the simulation with current observational constraints at $z>6$. Our two primary conclusions are:

\begin{itemize}

\item \bluetides\ predicts a strong relationship between the stellar mass and the far-UV escape fraction with a weaker relationship with observed UV luminosity. Galaxies with $M_{*}>10^{10}\,{\rm M_{\odot}}$ are predicted to have escape fractions around $0.1$ ($A_{\rm FUV}\approx 2$). This appears to be in tension with some observations of individual galaxies at $z>6$, however, due to the currently limited optical and sub-mm wavelength coverage observationally inferred stellar masses and escape fractions are uncertain. Assuming a IR SED which distributes more energy into the mid-IR (i.e. with a higher temperature or adding a mid-IR power-law component to the SED) can potentially increase the inferred obscured star formation rates by an order of magnitude bringing the observations in line with the predictions. Better characterisation of both the optical and IR SED is critical to overcome these issues.

\item Predictions for the surface density of dust reprocessed starlight powered sub-mm sources at $z>8$ lie well outside the range of sensitivities and areas probed by existing blank field surveys, and are thus consistent with a current lack of detected sources. Despite this, it is likely that ALMA will {\em eventually} build up sufficient area to either detect or place strong constraints on the numbers of obscured $z>8$ galaxies. 

\end{itemize}

\subsection*{Acknowledgements}

We acknowledge funding from NSF ACI-1036211 and NSF AST-1009781. The \bluetides\ simulation was run on facilities at the National Center for Supercomputing Applications. SMW acknowledge support from the UK Science and Technology Facilities Council through the Sussex Consolidated Grant (ST/L000652/1).

\bibliographystyle{mnras}
\bibliography{biblio} %

\appendix

\section{Data}

\begin{table*}
\caption{Tabulated values of the far-UV escape fraction as a function of stellar mass as used in Fig. \ref{fig:L_fesc}.}
\label{tab:L_fesc}
\begin{tabular}{cccccc}
\hline
   & \multicolumn{5}{c}{$\log_{10}(f_{\rm esc, FUV})=$} \\
$\log_{10}(M_*/{\rm M_{\odot}})$ & P$_{2.3}$ & P$_{15.9}$ & P$_{50}$ & P$_{84.1}$ & P$_{97.7}$ \\
\hline
\hline
$(8.0,8.25]$ & $-0.16$ & $-0.09$ & $-0.05$ & $-0.02$ & $-0.0$ \\
$(8.25,8.5]$ & $-0.26$ & $-0.17$ & $-0.1$ & $-0.04$ & $-0.01$ \\
$(8.5,8.75]$ & $-0.4$ & $-0.28$ & $-0.17$ & $-0.09$ & $-0.04$ \\
$(8.75,9.0]$ & $-0.56$ & $-0.42$ & $-0.28$ & $-0.16$ & $-0.07$ \\
$(9.0,9.25]$ & $-0.73$ & $-0.58$ & $-0.42$ & $-0.25$ & $-0.13$ \\
$(9.25,9.5]$ & $-0.89$ & $-0.73$ & $-0.57$ & $-0.37$ & $-0.22$ \\
$(9.5,9.75]$ & $-1.04$ & $-0.88$ & $-0.7$ & $-0.48$ & $-0.29$ \\
$(9.75,10.0]$ & $-1.14$ & $-0.98$ & $-0.82$ & $-0.61$ & $-0.39$ \\
$(10.0,10.25]$ & $-1.24$ & $-1.08$ & $-0.91$ & $-0.72$ & $-0.44$ \\
$(10.25,10.5]$ & $-1.28$ & $-1.21$ & $-1.09$ & $-0.88$ & $-0.79$ \\
$(10.5,10.75]$ & $-1.44$ & $-1.37$ & $-1.19$ & $-1.13$ & $-1.09$ \\
\hline
\end{tabular}
\end{table*}

\begin{table*}
\caption{Tabulated values of the far-UV escape fraction as a function of the observed UV luminosity as used in Fig. \ref{fig:L_fesc_L1500}.}
\label{tab:L_fesc_L1500}
\begin{tabular}{cccccc}
\hline
   & \multicolumn{5}{c}{$\log_{10}(f_{\rm esc, FUV})=$} \\
$\log_{10}(L_{\rm FUV}^{\rm obs}/{\rm erg\, s^{-1}\, Hz^{-1}})$ & P$_{2.3}$ & P$_{15.9}$ & P$_{50}$ & P$_{84.1}$ & P$_{97.7}$ \\
\hline
\hline
$(27.875,28.125]$ & $-0.2$ & $-0.11$ & $-0.06$ & $-0.02$ & $-0.0$ \\
$(28.125,28.375]$ & $-0.32$ & $-0.16$ & $-0.07$ & $-0.02$ & $-0.0$ \\
$(28.375,28.625]$ & $-0.57$ & $-0.28$ & $-0.12$ & $-0.04$ & $-0.01$ \\
$(28.625,28.875]$ & $-0.77$ & $-0.44$ & $-0.2$ & $-0.08$ & $-0.02$ \\
$(28.875,29.125]$ & $-0.93$ & $-0.57$ & $-0.29$ & $-0.13$ & $-0.05$ \\
$(29.125,29.375]$ & $-1.02$ & $-0.69$ & $-0.38$ & $-0.19$ & $-0.08$ \\
$(29.375,29.625]$ & $-1.09$ & $-0.82$ & $-0.49$ & $-0.27$ & $-0.16$ \\
$(29.625,29.875]$ & $-0.85$ & $-0.74$ & $-0.46$ & $-0.42$ & $-0.26$ \\
\hline
\end{tabular}
\end{table*}

\end{document}